# A New Model For Incorporating
# the Varistions of the Dielectric Permittivity
# of the Medium: Part (1)
# Potential Distribuition of an Electrical
# Double Layer in Rectangular Nanochannel by Applying
# Finite Element Method Technique.

**Rajendra Padidhapu[1], Shahnaz Bathul[2], V.Brahmajirao[3]**

[1] Department of Mathematics, Flora Institute of Technology, 49/1, Khopi, Pune, Maharastra, India. [2] Department of Mathematics, JNTUCEH, Jawaharlal Nehru Technological University, Hyderabad, A.P., India. [3] School of Biotechnology, MGNIRSA, A Unit of Swaminathan Research Foundation, Hyderabad, A.P., India. E-mail: rajendra.padidhapu@gmail.com

### ABSTRACT

The effect of the dielectric permittivity of the electrolytic medium in the Electrical Double Layer (EDL) in Nanochannel is an important factor that requires a correction, before the application of the model in various chemical and biomedical examinations. However very few efforts are put forth to accommodate this into the mathematics of the EDL Models developed hitherto, even though it was realized by a host of authors for the past half a century or even more. Most of the Nanochannel made lab-on-chips in a rectangular cross section require suitable impregnation of this. The effect of the EDL in a nanochannel isimportant in various chemical and biomedical examinations. Most of the nanochannelmadelab-on-chips in a rectangular cross section. The EDL insuch a nanochannel is governed by two dimensional Gouy-Chapman (GC)equation. In this paper, the 2-dimensionalGCequation with the variation of dielectric permittivity is solved in two ways:(i)Analytically solution by separation of variable method, (ii) Numerical solution by finite element method. The potential distribution calculated by using these two different solutions are compared and discussed. The analytic and numerical solutions show the important impacts of the nanochannel.

# 1. Preliminary prescription for the formation of the system and analysis of the model:

## 1.1 Choice of ingredients and their location, Nanochannel and transport of ionic matter:

Micro channels are broadly defined as channels whose dimensions are less than 1 millimeter and greater than 1 micron. Nano channels have characteristic dimensions anywhere from the submicron scale i.e., a fraction of that of the micro channel. The flow in such systems exhibits behavior that is the same as in most macroscopic systems (K.V.Sharp et.al., 2005) and they offer advantages due to their high surface-to-volume ratio and their small volumes. These systems find application in the area of MEMS devices for biological and chemical analysis. These tools transport biological ingredients such as proteins, DNA, cells, and embryos and are also helpful to transport chemical samples and analysts.

## 1.2 Forces and Interactions possible:

Janson et al.,(1999), Gad-el-Hak, (1999), considering the electro kinetic effect observe that the effects of molecular structure are quite different in gases and liquids. If the Knudsen number (defined as, $K_n = \left[\frac{\lambda}{L_s}\right]$ where $\lambda$ is the mean free path in a gas and $L_s$ is the characteristic channel dimension which is greater than $10^{-3}$. Conditions of Non-equilibrium are supposed to occur if it is greater than **.** Fluids that are Newtonian at ordinary rates of shear and extension can become non-Newtonian at very high rates. The pressure gradient becomes especially large in small cross section channels. For fixed volume flux, the pressure gradient increases to.Chemical interaction causes the Electro kinetic effects that appear at the interface between liquids and solids such as glass (2005).Consequently the formation of an electrically charged double layer that induces a charge distribution in a very thin layer of fluid close to the wall comes to existence. Hence an electric field acts cross this layer,(1999) it creates a body force capable of moving the fluid as if it were slipping over the wall. Analysis of such flow requires consideration of different physical phenomena. Arkilic, E.B et.al., (1997). and Harley, J.C.,et,al., (1995) explore in detail about this. Bridgman in 1923 , evaluated the lattice spacing ,and Probstein (1994) modified the treatment to obtain an expression for the same .

## 1.3 Need to account for the variations in the dielectric constant of the aqueous binary solvent electrolytic system:

P.G.Wolyness (1980)  in an interesting review on the Dynamics of Electrolytic solutions affirms that the accurate measurements possible and consequent findings around the turn of the century ,underline the need for the theoreticians to develop appropriate models that can peep into the long range interactions to a better approximation. D.B.Yaakov et.al (2010), proposing a mean-field model to account for the heterogeneity of the dielectric constant caused by the ions suggested that the effect of ions on the local dielectric constant should be taken into account when interpreting experiments that address ion-specific effects. Many theoretical studies that were reported in literature in the past half a century (Pitzer K.S. 1993), attribute these findings to ion-specific effects. While trying to interpret a large bulk of experimental evidence, and consequent interesting conclusions by a host of authors (Sastry P.S. 1970; Rao P.S.K.M. 1967; Glueckauff S. 2007; Hitoshi Ohtaki 1993) in respect of Dielectric constant (Sastry P.S. 1970; Rao P.S.K.M. 1967; Marcus Y. 1997) Surface Tension & Refractive index (Raja Rao N.S.S.V. 2013) , and interionic force related measurements of thermodynamic parameters ( Brahmajirao V. 1977; Wolyness 1980) like ,Entropy, Gibbs free energy, Activity Coefficient (Hitoshi Ohtaki 1993) ,Osmotic coefficient , Ultrasonic velocity, Density (Meng P.F. 2007) , Viscosity ,Adiabatic&Apparent Molar Compressibility (Akhilan C. 2008; Onuki et. al., 2012; Raja Rao N.S.S.V. 2013) Apparent molal volumes ,Excess parameters, Acoustic Impudence etc., and their variations in electrolytic solutions in binary aqueous solvent mixtures (Raja Rao N.S.S.V. 2013) ,as functions of concentration and temperature ,making use of most sophisticated techniques between charged ionic systems , established the existence of several new mechanisms involving solute-solvent-water interactions to explain the results, This could give a firm and conclusive evidence to the Nobel Prize winning Eigen &Tamm Mechanism of ion pair formation (Eigen M. 1954) in electrolytic ionic liquid systems involving the stepwise [Akhilan C. 2008;Brahmajirao V. 1977) physiochemical interaction (Akhilan C. 2008; Raja Rao N.S.S.V. 2013) forming three types of ion pairs namely (1) solvent separated, (2)solvent shared & (3) contact ion pairs. Dispersion forces and specific surface-ion interactions [34] play a prominent role in this mechanism.These experimental findings suggest another source of ion-specificity originating from the local variations of the dielectric constant due to the presence of ions in the solution.

Glueckauff's (Glueckauff S. 2007) using his model developed a detailed expression for the depression of the dielectric constant ,of the dielectric continuum almost Fifty years back .This forms the backdrop of the work reported in the paper

## 2. The new model incorporating the dielectric depression:

Electro-osmotic flowis the bulk fluid motion that results when an externally applied electric field interacts with the net surplus of charged ions in the diffuse part of an electrical double layer (EDL). EDL is a very thin region of non-zero net charge density near solid-liquid interface and thisis an important parameterin nanochannels, since theyhave

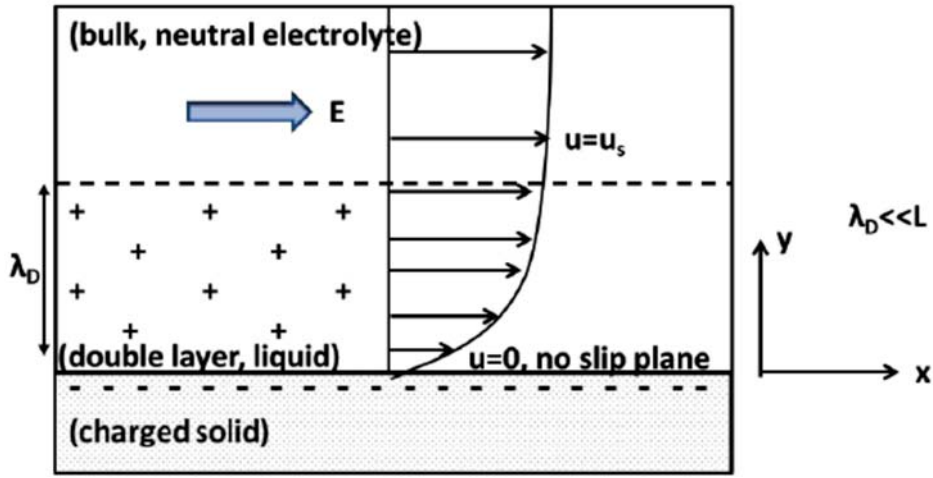

**Fig 1:**Schematic of an Electro-osmotic flow

The ions within the double layer are very close to the wall (see Fig. 1) that gives a large electrostatic force, while those ions in the bulk fluid feel a relatively small electrostatic force. The potential drop across the double layer is defined as the difference in potential between the bulk fluid and the no-slip plane which is known as zeta potential $(\zeta)$. The calculation of the $\zeta$ depends on the EDL models.

The EDL was initially represented by Helmholtz with a simple capacitor. Gouy and Chapman treated one layer of charge (Overbeek, J.T.G. 1982) The most common representation of the EDL was Stern model (Fig.2) that is separated into three layers (Hunter R.J. 1981).The 1st layerprovides the potential $\psi_i$, where co-ions and counterions are not hydrated and are specifically adsorbed to the surface. The 2nd layer withpotential$\psi_d$, consisting of hydrated, and partially hydrated counter ions. The 3rd layer is the diffuse layer, composed of mobile co-ions and counter ions, in which resides the slip plane providing the $\zeta$ potential. The diffusion layer and the slip plane are located close to each other (Bhatt K. 2005) allowing the approximationof, *with* the potential.

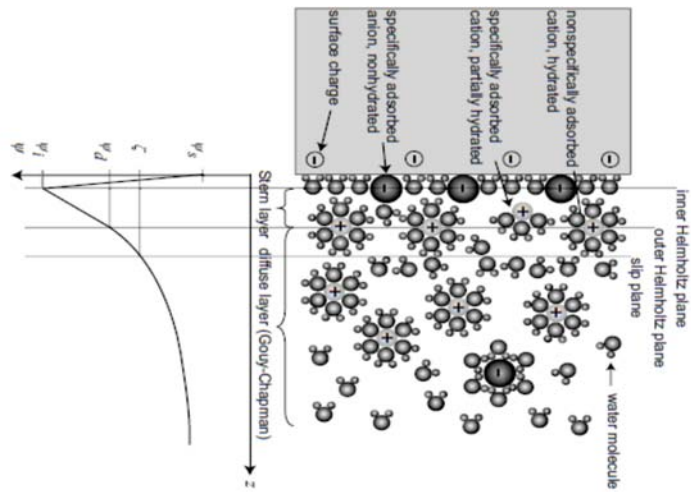

**Fig 2:** Gouy-Chapman-Stern EDL model of the solid-electrolyte interface.

3. Mathematical Modeling of Electrical Double Layer (EDL) in Rectangular Nanochannel:
Consider a rectangular nanochannel of length L, width W and height H as shown in Fig 3
formed between two parallel silicon plates containing an aqueous NaCl electrolyte solution
under an external voltage bias.Since the fluidic channel sizes used in today's investigation
have already advanced $\sim 3nm$ (Sparreboom A. 2010) all through this paper it is supposed
that the continuum hypothesis is valid. If the Electrical Double Layer can develop fully in
the nanochannel, the electric potential reduces to zero in the center. The space-charge
model(see Gross R. et. al., 1968; Wang X.L. 1995) gives complete physical interpretations
of ion transport through nanometer-sized channels. The space-charge model has been
simplified by (Pennathur and Santiago 2005) to describe electro-kinetic transport in long
nanochannels, confirmed by their experimental studies (Pennathur and Santiago 2005).

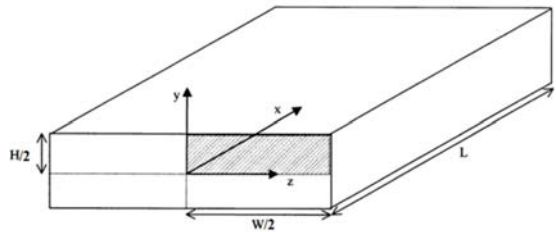

**Fig 3:** Computational domain &! of length 'L', width $\frac{W}{2}$ and height $\frac{H}{2}$ of nanochannel
formed between two parallel silicon plates containing an aqueous electrolyte solution
under an external voltage bias.

The electro-chemical potential $\bar{\mu}$ of ion i nearer to charged surface in a liquid phase at constant pressure, P and temperature is defined (Israelachvili J. 1992) as

$$\bar{\mu} = \mu + z_i F \psi \qquad (1)$$

where $\mu$ is the chemical potential, F is the Faraday constant, $\psi$ is the electric potential due to the surface charge. Therefore by, electrical potential and net charge density per unit volume, at any point in the aqueous electrolyte solution is given by Poisson equation

$$\nabla^2 \psi = -\frac{\rho}{\varepsilon_0 \varepsilon_r} \qquad (2)$$

where $\varepsilon_0$ is the permittivity of the vacuum, $\varepsilon_r$ is the dielectric constant which is varying smoothly and continuously with respect to concentration inside the compact layer of aqueous electrolyte solution (see (Yang X. et. al., 2007; Yang X. et. al., 2008; Rajendra Padidhapu et. al., 2013) and it is given by

$$\varepsilon_r = \varepsilon + \beta c \qquad (2a)$$

where $\varepsilon_r$ is the dielectric constant of the aqueous electrolyte solution, $\varepsilon = 80$ is the dielectric constant of pure water and $\beta$ is a phenomenological coefficient (in units of M-1) of the linear term. The most common case where $\beta < 0$ corresponds to a dielectric decrement. Combined with Nernst-Plank equation governed by
governed by

$$\frac{\partial c_i}{\partial t} = \nabla \left( D_i \nabla c_i + \frac{z_i F D_i c_i \nabla \psi}{RT} \right) \qquad (3)$$

where $D_i, z_i$ and $c_i$ are the diffusivity, the charge valance and the concentration of the ionic species i respectively, t is time, F is the Faraday constant, R is the gas constant, and T is the absolute temperature.

$\mu$ is also known as Gibbs Energy and it is defined as

$$\mu = \mu_i^0 + RT ln\left(\frac{c_i}{c_i^\infty}\right) \qquad (4)$$

where $\mu_i^0$ is the standard state of chemical potential of ion i at constant temperature T and Pressure P.

$RT ln\left(\frac{c_i}{c_i^\infty}\right)$ is the Gibbs Free Energy which is deviating from ideality.

Substitute (4) into (1) gives the electro-chemical potential $\bar{\mu}$:

potential $\bar{\mu}$:

$$\bar{\mu} = \mu_i^0 + RTln\left(\frac{c_i}{c_i^\infty}\right) + z_i F\psi \qquad (5)$$

The electro-chemical potentials $\bar{\mu}$ must be constant at thermodynamic equilibrium; otherwise, the system would be able to reduce its Gibbs free energy.

$$\therefore \nabla(\bar{\mu}) = 0 \qquad (6)$$

Equation (6) becomes

$$RT\nabla ln\left(\frac{c_i}{c_i^\infty}\right) = -z_i F\nabla\psi \qquad (7)$$

We have the relation between Faradays constant and Boltzmann constant is given by

$$\frac{e}{k_B} = \frac{F}{R} \qquad (8)$$

Where e is the elementary charge, R is the gas constant.
Substitute in equation (7), which gives

$$k_B T\nabla ln\left(\frac{c_i}{c_i^\infty}\right) = -z_i e\nabla\psi \qquad (9)$$

Consider two infinite parallel charged surfaces symmetrically placed at y= +H/2 and y= "H/2respectively. Also consider an aqueous electrolyte solution havingNions each of concentration $c_i$and valence $Z_i$ in between the charged surface.To solve the equation (9), approximate the surface as aninfinite planar plate which resides in the half-space ofy<0. The potential at the surface isknown as the æ potential. At infinity, $\psi(") = 0$, and bulk concentration $c_i^\infty = c_0$.Apply the above boundary conditions and integrate the equation (9) to get theionic concentrations

$$c_i = c_0\, exp\left[-\frac{z_i e}{k_B T}\psi\right] \qquad (10)$$

The charge density for N ions is defined as

$$\rho = ze\sum_{i=0}^{N} z_i c_i \qquad (11)$$

Substitute equation (10), (11) in equation (2) gives

$$\nabla^2\psi = -\frac{e}{\varepsilon_0\varepsilon_r}\sum_{i=0}^{N} z_i c_i exp\left[-\frac{z_i e}{k_B T}\psi\right] \qquad (12)$$

which is known as Poisson-Boltzmann equation. This Poisson- Boltzmann theory neglects ion specific effects, which can be taken into account only when depression in dielectric constant is accounted for (Wolyness P.G. 1980)

The charge density for symmetric ions is defined as

$$\rho = -2ze[c_\pm] = -2zec_0 exp\left[-\frac{ze}{k_BT}\psi\right] = -2zec_0 sinh\left[\frac{ze}{k_BT}\psi\right] \qquad (13)$$

Therefore for symmetric ions, The PB equation (12) reduces to the Gouy-Chapman (GC) equation [16].

$$\nabla^2\psi = \frac{2ze^2c_0}{\varepsilon_0\varepsilon_r} sinh\left[\frac{ze}{k_BT}\psi\right] \qquad (14)$$

Assume that the nanochannel has rectangular cross-section, the fluid flow is directed along the x-axis and yz is the cross section plane, then the GC equation (14) becomes

$$\frac{\partial^2\psi}{\partial y^2} + \frac{\partial^2\psi}{\partial z^2} = \frac{2ze^2c_0}{\varepsilon_0\varepsilon_r} sinh\left[\frac{ze}{k_BT}\psi\right] \qquad (15)$$

The dimensionless variables are introduced as follows:

$$Y = \frac{y}{D_h}, Z = \frac{z}{D_h}, \Psi = \frac{ze}{k_BT}\psi \qquad (16)$$

where Y is the non-dimensional y axis, Z is the non-dimensional z axis, $\Psi$ is the non-dimensional surface potential and $D_h$ is the hydraulic diameter of the nanochannel which is given by the formula

$$D_h = \frac{2HW}{H+W} \qquad (17)$$

Introduce Debye Length $k^{-1}$, the inverse of the characteristic EDL thickness

$$k^{-1} = \sqrt{\frac{\varepsilon_0\varepsilon_r k_BT}{2ze^2c_0}} \qquad (18)$$

Substitute equation (16), (17), (18) in equation (14) gives

$$\frac{\partial^2\Psi}{\partial Y^2} + \frac{\partial^2\Psi}{\partial Z^2} = (kD_h)^2 sinh(\Psi) \qquad (19)$$

with the boundary conditions

$$Y = 0, \qquad \frac{\partial\Psi}{\partial y} = 0 \qquad (19a)$$

$$Z = 0, \qquad \frac{\partial \Psi}{\partial z} = 0 \tag{19b}$$

$$Y = \frac{H}{2}, \quad \Psi = \frac{ze}{k_B T}\zeta \tag{19c}$$

$$Z = \frac{W}{2}, \qquad \Psi = \frac{ze}{k_B T}\zeta \tag{19d}$$

where $\zeta$ is the electric potential at the surface.

In practice, solid-liquid of nanochannel surfaces have a surface electrical potential $\Psi$ is small comparedwith the thermal energy of ions; i.e., $|ze\Psi| < k_B T$. Therefore, the Gouy-Chapman (GC) equation can be linear approximated by first terms in the Taylor series. This transforms equation (19) to:Taylor series.

$$\frac{\partial^2 \Psi}{\partial Y^2} + \frac{\partial^2 \Psi}{\partial Z^2} = (kD_h)^2 \Psi \tag{20}$$

with the boundary conditions

$$Y = 0, \qquad \frac{\partial \Psi}{\partial y} = 0 \tag{20a}$$

$$Z = 0, \qquad \frac{\partial \Psi}{\partial z} = 0 \tag{20b}$$

$$Y = \frac{H}{2}, \quad \Psi = \frac{ze}{k_B T}\zeta \tag{20c}$$

$$Z = \frac{W}{2}, \qquad \Psi = \frac{ze}{k_B T}\zeta \tag{20d}$$

where $\zeta$ is the electric potential at the surface.

### 3. Analytic Solution by Separation of Variable Method:

By using the separation of variable method, the solution to the linearized Gouy-Chapman (GC)equation (20) can be obtained.The general solution of the equation (20) can be written as

$$\Psi(Y, Z) = \Psi_c(Y, Z) + \Psi_p(Y, Z) \tag{21}$$

where $\Psi_c(Y, Z)$ the solution of

$$\frac{\partial^2 \Psi}{\partial Y^2} + \frac{\partial^2 \Psi}{\partial Z^2} = 0 \tag{22}$$

which is called complementary function and $\Psi_p(Y,Z)$ is a function of (Y,Z) that satisfies the equation (20) which is called particular integral. First, we derive complementary solution and then particular integral to get complete solution of equation (20).

Let $\Psi_c(Y,Z) = UV$, where U is the functions of X alone and V is the functions of Y alone be a solution of equation (21), then

$$\frac{\partial^2 \Psi}{\partial Y^2} = U''V \qquad \text{and} \qquad \frac{\partial^2 \Psi}{\partial Z^2} = UV''$$

. Substituting these values in equation (1), we get

$$U''V + UV'' = 0$$

$$\Rightarrow \frac{U''}{U} = -\frac{V''}{V} \tag{23}$$

As a L.H.S is being a function of X alone and R.H.S is function of Y alone, and X and y are being independent variables, equation (23) will holds good only if both side reduces to a constant, say 'c'

Therefore, equation (23) leads to

$$\frac{U''}{U} = c \ \ and \ -\frac{V''}{V} = c$$

$$or \ \ U'' - cU = 0 \ and \ V'' + cV = 0$$

$$\therefore (D^2 - c)U = 0 \ and \ (D'^2 + c)V = 0 \tag{24}$$

Where $D$ $stands$ $for$ $\frac{\partial^2}{\partial U^2}$ and $D'$ $stands$ $for$ $\frac{\partial^2}{\partial V^2}$

The complementary solution of this physical system is

$$\Psi_c(Y,Z) = (c_1 \cosh mY + c_2 \sinh mY)(c_3 \cosh mZ + c_4 \sinh mZ) \tag{25}$$

where $c_1, c_2, c_3$ and $c_4$ are arbitrary constants

The particular integral of the equation (20) is

$$\Psi_p(Y,Z) = \frac{K}{(D^2 - c)(D'^2 + c)} = -\frac{K}{4c^2} \tag{26}$$

where K is a constant, i.e. $K = (kD_h)^2 \frac{ze}{k_B T}\zeta$

Therefore, from equation (25) and (26), the most general solution of equation (20) is

$$\Psi(Y,Z) = (c_1 \cosh mY + c_2 \sinh mY)(c_3 \cosh mZ + c_4 \sinh mZ) - \frac{K}{4c^2} \qquad (27)$$

Apply boundary condition of equation (20a), (20b), (20c), (20d) on equation (27) that gives arbitrary constants $c_1$, $c_2$, $c_3$ and $c_4$.

Therefore, the complete analytical solution of electrical potential distribution in $\frac{1}{4}$ of the rectangular nanochannel is of the form:

$$\Psi(Y,Z) = 4\frac{ze}{k_B T}\zeta \left[ \sum_{m=1}^{\infty} \frac{(-1)^{m+1}\cosh\left[\sqrt{1+\frac{(2m-1)^2\pi^2 D_h^2}{k^2 W^2}}kY\right]}{(2m-1)\pi\cosh\left[\sqrt{1+\frac{(2m-1)^2\pi^2 D_h^2}{k^2 W^2}}\frac{kH}{2D_h}\right]} \cos\frac{(2m-1)\pi D_h}{W}Z \right.$$
$$\left. + \sum_{n=1}^{\infty} \frac{(-1)^{n+1}\cosh\left[\sqrt{1+\frac{(2n-1)^2\pi^2 D_h^2}{k^2 H^2}}kZ\right]}{(2n-1)\pi\cosh\left[\sqrt{1+\frac{(2n-1)^2\pi^2 D_h^2}{k^2 H^2}}\frac{kW}{2D_h}\right]} \cos\frac{(2n-1)\pi D_h}{H}Y \right] \qquad (28)$$

## 4. Numerical Solution by Finite Element Method (FEM):

We can solve the equation (20) numerically in two ways, one is Finite Difference Method (FDM) and the other is Finite element method (FEM). FDM represents the solution region by array of grid points; its application becomes problematic with problems having irregularly shaped boundaries .Such problems can be handled more easily by using the FEM. In FDM, it is possible to write down the algebraic equations directly from PDE. This is not possible in FEM where supplementary steps are required.We theoretically break down the FEM in four parts ( Fish J. et. al., 2007) :the strong form, the weak form, approximation of functions, and the discrete equations.

In order to find the electric potential for the system of equations (20),(20a), (20b), (20c) and (20d)by Finite Element Method, the variational formation is constructed over a typical element. First, we multiply the equation (20) with the test function V(Y,Z) and integrate the result over a typical element $R_e$ gives ....................................................................

$$........ \iint_{R_e} \left[ \left( \frac{\partial \Psi}{\partial Y}\frac{\partial V}{\partial Y} + \frac{\partial \Psi}{\partial Z}\frac{\partial V}{\partial Z} \right) - VC \right] dY dZ - \int_{c_e} V q_n \, ds = 0 \qquad (29)$$

where $q_n = \eta_Y \frac{\partial \Psi}{\partial Y} + \eta_Z \frac{\partial \Psi}{\partial Z}$, $\eta_Y$ and $\eta_Z$ directional cosines of a unit vector $\hat{n}$ on the boundary

$c_\varepsilon$ and ds is

an arc lengthof infinitesimal element along the boundary.

To get the finite element of the given equation, we approximate $\Psi$ by the expression

$$..... \quad \Psi = \sum_{j=1}^{n} u_j \phi_j \tag{30}$$

Where $\Psi_j = \Psi(Y_j, Z_i)$ and $\Phi_j$ have the property

$$\phi_i((Y_j, Z_j) = \delta_{ij} = \begin{cases} 1, & if \ i = j \\ 0, & if \ i \neq j \end{cases} \tag{31}$$

Substituting equation (30) in equation (29), and putting $V = \phi_i$, we obtain

$$\sum_{j=1}^{n} \iint_{R_\varepsilon} \left( \frac{\partial \phi_i}{\partial Y} \frac{\partial \phi_j}{\partial Y} + \frac{\partial \phi_i}{\partial Z} \frac{\partial \phi_j}{\partial Z} \right) \Psi_j dY dZ - \iint_{R_\varepsilon} C \phi_i \, dY dZ - \int_{c_\varepsilon} \phi_i q_n \, ds = 0 \ \ for \ i = 1,2,....n \tag{32}$$

The equation (320 can be written in the form

$$\sum_{j=1}^{n} K_{ij}^{(\varepsilon)} \Psi_j^{(\varepsilon)} = F_i^{(\varepsilon)} \tag{33}$$

Where

$$K_{ij}^{(\varepsilon)} = \iint_{R_\varepsilon} \left( \frac{\partial \phi_i}{\partial Y} \frac{\partial \phi_j}{\partial Y} + \frac{\partial \phi_i}{\partial Z} \frac{\partial \phi_j}{\partial Z} \right) dY dZ \tag{34}$$

and

$$F_i^{(\varepsilon)} = \iint_{R_\varepsilon} C \phi_i \, dY dZ + \int_{c_\varepsilon} \phi_i q_n \, ds \tag{35}$$

Equation (33) represents the finite element model of the linearized Gouy-Chapman (GC) equation (20).

We next consider the triangular element in which nodes are numbered in the counter-clockwise direction and derive the interpolation functions for it. We assume the interpolating polynomial in such a way that the number of terms in it equals the number of nodes in the triangular element. Accordingly, we assume

$$\Psi(Y, Z) = a_1 + a_2 Y + a_3 Z \tag{36}$$

as the required approximation. We also set where $(Y_i, Z_i), i = 1,2,3$ denotes the three vertices of the triangle. Substituting equation (37) in equation

(36), we obtain

$$\Psi_1 = a_1 + a_2 Y_1 + a_3 Z_1$$
$$\Psi_2 = a_1 + a_2 Y_2 + a_3 Z_2$$
$$(38)$$
$$\Psi_3 = a_1 + a_2 Y_3 + a_3 Z_3$$

Solving equation (38), we obtain

$$a_1 = \frac{1}{2\Delta_e} \begin{vmatrix} \Psi_1 & \Psi_2 & \Psi_3 \\ Y_1 & Y_2 & Y_3 \\ Z_1 & Z_2 & Z_3 \end{vmatrix}$$

$$a_2 = \frac{1}{2\Delta_e} \begin{vmatrix} \Psi_1 & \Psi_2 & \Psi_3 \\ Z_1 & Z_2 & Z_3 \\ 1 & 1 & 1 \end{vmatrix} \tag{39}$$

$$a_1 = \frac{1}{2\Delta_e} \begin{vmatrix} \Psi_1 & \Psi_2 & \Psi_3 \\ 1 & 1 & 1 \\ Y_1 & Y_2 & Y_3 \end{vmatrix}$$

where

$$\Delta_e = Area \ of \ the \ triangle = \frac{1}{2} \begin{vmatrix} 1 & Y_1 & Z_1 \\ 1 & Y_2 & Z_2 \\ 1 & Y_3 & Z_3 \end{vmatrix} \tag{40}$$

Substituting $a_1, \ a_2, \ a_3$ in equation (36) and simplifying, we obtain

$$\Psi(Y, Z) = \frac{1}{2\Delta_e} [\Psi_1 (Y_2 Z_3 - Y_3 Z_2) + \Psi_2 (Y_3 Z_1 - Y_1 Z_3) + \Psi_3 (Y_1 Z_2 - Y_2 Z_1)]$$
$$+ \frac{1}{2\Delta_e} [\Psi_1 (Z_2 - Z_3) + \Psi_2 (Z_3 - Z_1) + \Psi_3 (Z_1 - Z_2)]$$
$$+ \frac{1}{2\Delta_e} [\Psi_1 (Y_3 - Y_2) + \Psi_2 (Y_1 - Y_3) + \Psi_3 (Y_2 - Y_1)] \tag{41}$$

Collecting the coefficients of $\Psi_1, \Psi_2 \ and \ \Psi_3$ in the above equation (41) can be written in the form

Where $\phi_j^{(e)}$ are the linear interpolation functions for the triangular elements under consideration, and are given by

$$\phi_1^{(e)}(Y,Z) = \frac{1}{2\Delta_e} \begin{vmatrix} 1 & Y & Z \\ 1 & Y_2 & Z_2 \\ 1 & Y_3 & Z_3 \end{vmatrix}$$

$$\phi_2^{(e)}(Y,Z) = \frac{1}{2\Delta_e} \begin{vmatrix} 1 & Y & Z \\ 1 & Y_3 & Z_3 \\ 1 & Y_1 & Z_1 \end{vmatrix} \tag{43}$$

$$\phi_3^{(e)}(Y,Z) = \frac{1}{2\Delta_e} \begin{vmatrix} 1 & Y & Z \\ 1 & Y_1 & Z_1 \\ 1 & Y_2 & Z_2 \end{vmatrix}$$

From the formulae (70), it is easily verified that

$$\phi_i^{(e)}((Y_j,Z_j)) = \begin{cases} 1, & if\ i = j \\ 0, & if\ i \neq j \end{cases} \quad and \quad \sum_{i=1}^{3} \phi_i^{(e)}(Y,Z) = 1 \tag{44}$$

We also have

$$\frac{\partial \phi_i^{(e)}}{\partial Y} = \frac{Z_2 - Z_3}{2\Delta_e}, \frac{\partial \phi_i^{(e)}}{\partial Z} = \frac{Y_3 - Y_2}{2\Delta_e}$$

$$\frac{\partial \phi_i^{(e)}}{\partial Y} = \frac{Z_3 - Z_1}{2\Delta_e}, \frac{\partial \phi_i^{(e)}}{\partial Z} = \frac{Y_1 - Y_3}{2\Delta_e} \tag{45}$$

$$\frac{\partial \phi_i^{(e)}}{\partial Y} = \frac{Z_1 - Z_2}{2\Delta_e}, \frac{\partial \phi_i^{(e)}}{\partial Z} = \frac{Y_2 - Y_1}{2\Delta_e}$$

Using equation (45) the element matrices $K_{ij}^{(e)}$ and $F_i^{(e)}$ in equation (33) and we can computed easily. If the number of elements is increased, then the accuracy of the finite element solution can be improved.

**5. C** solu the para para elec 5 in

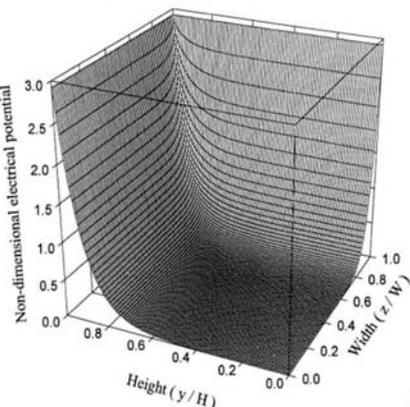

Fig 4(a)

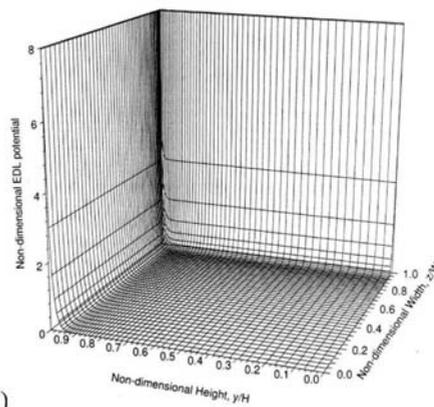

Fig 4(b)

**Fig 4:** Non dimensional electrical potential profile in 1/4$^{th}$ rectangular nanochannel.

Certainly, the linear approximation for solving the Gouy-Chapman equation is valid only when Zeta potential is small (Usually less than 25mV is suggested). However if one wants to study the flow behavior of highly charged system where only the outer region of the diffusion layer is important, the linear approximation gives a good prediction even when the zeta potential is upto100mV. The analytical solution of Gouy-Chapman (GC) equation (28) ispresented in Fig 4(a). The numerical solution of Gouy-Chapman (GC) equation (33) is presented in Fig 4(b).